\documentclass[lettersize,journal]{IEEEtran}

\usepackage{graphicx} 
\usepackage{subfigure} 
\usepackage{amsmath,amssymb,amsfonts,mathrsfs,bm}
\usepackage{algorithm,algorithmic}
\providecommand\algocomment[1]{ {\footnotesize // #1 } }
\usepackage{multirow}
\usepackage{textcomp}
\usepackage{xcolor}
\usepackage{cite}
\usepackage{booktabs}
\usepackage{array}
\usepackage{caption}
\usepackage{multicol}
 \usepackage{threeparttable}

\allowdisplaybreaks
\usepackage[hidelinks, breaklinks = true]{hyperref}  

\newtheorem{remark}{Remark}
\newcommand{\signbin}{\texttt{sign\_to\_bin}}
\newcommand{\sign}{\texttt{sign}}
\newcommand{\binsign}{\texttt{bin\_to\_sign}}
\newcommand{\syn}{\texttt{syn}}

\begin{document}

\title{Separate Source Channel Coding Is Still What You Need: An LLM-based Rethinking}

\author{Tianqi Ren, Rongpeng Li, Ming-min Zhao, Xianfu Chen, Guangyi Liu, Yang Yang, Zhifeng Zhao and Honggang Zhang \\
\thanks{T. Ren, R. Li and M. Zhao are with College of Information Science and Electronic Engineering, Zhejiang University, Hangzhou 310027, China. X. Chen is with Shenzhen CyberAray Network Technology Co., Ltd, Shenzhen 518000, China. G. Liu is with China Mobile Research Institute, Beijing 100053, China. Y. Yang is with The Internet of Things Thrust, The Hong Kong University of Science and Technology, Guangzhou 511453, China. Z. Zhao is with Zhejiang Lab, Hangzhou 311121, China. H. Zhang is with Faculty of Data Science, The City University of Macau, Macau 999078, China.}
\thanks{Corresponding Author: Rongpeng Li (lirongpeng@zju.edu.cn).}

}
\maketitle

\begin{abstract}
Along with the proliferating research interest in Semantic Communication (SemCom), Joint Source Channel Coding (JSCC) has dominated the attention due to the widely assumed existence in efficiently delivering information semantics. 
Nevertheless, this paper challenges the conventional JSCC paradigm and advocates for adopting of Separate Source Channel Coding (SSCC) to enjoy the underlying more degree of freedom for optimization. 
We demonstrate that SSCC, after  leveraging the strengths of Large Language Model (LLM) for source coding and Error Correction Code Transformer (ECCT) complemented for channel decoding, offers superior performance over JSCC. 
Our proposed framework, effectively highlights the compatibility challenges between SemCom approaches and digital communication systems, particularly concerning the resource costs associated with the transmission of high precision floating point numbers. 
Through comprehensive evaluations, we establish that assisted by LLM-based compression and ECCT-enhanced error correction, SSCC remains a viable and effective solution for modern communication systems. In other words, separate source channel coding is still what we need!
\end{abstract}


\begin{figure*}[tb]
    \centering
    \includegraphics[width=.85\textwidth]{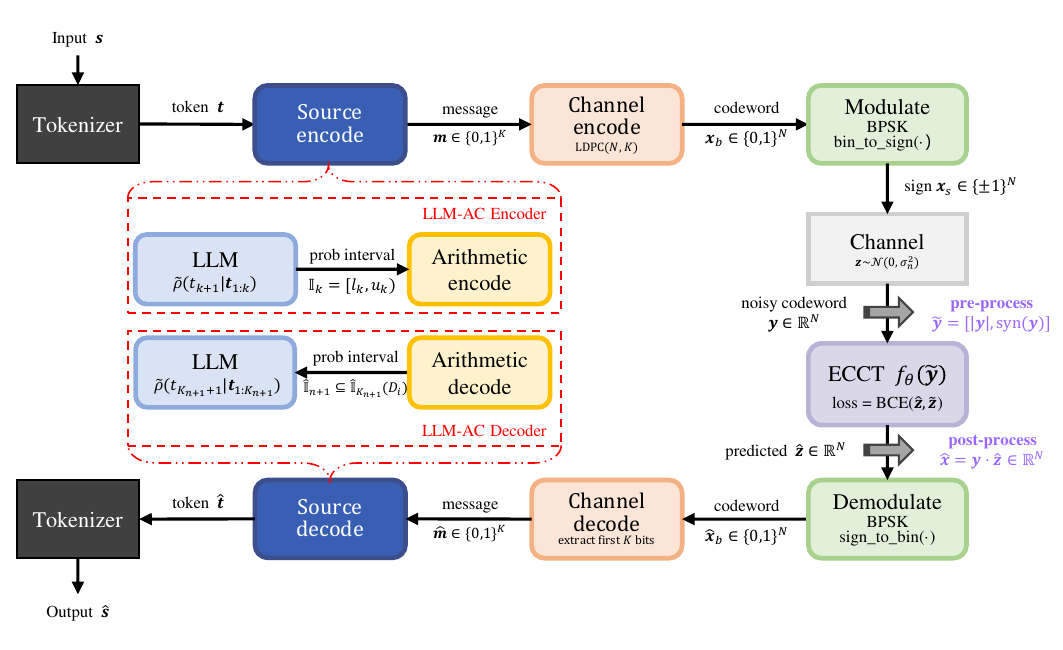}
    \caption{Framework of LLM-based and ECCT-complemented SSCC system.}
    \label{fig:System_Model}
\end{figure*}


\section{Introduction}
\label{sec:intro}
    Semantic communication (SemCom) has garnered significant attention in recent years, with researchers exploring innovative approaches to enhance the efficiency and reliability of information transmission \cite{lu_semanticsempowered_2024}. Generally, SemCom leverages deep learning-based Joint Source-Channel Coding (JSCC) methods to preserve global semantic information and local texture during the transmission process. 
    DeepJSCC \cite{kurka2020deepjscc} pioneers these works by implementing JSCC with feedback and allowing for real-time adaptation to channel conditions. 
    Along with its steady progress, JSCC has been substantially studied, mostly with the optimization objective shifting from bit error rates to the semantic relevance of the transmitted information in SemCom \cite{xie2021deep,tong2022image, tong_alternate_2023,tong_alternate_2024, zhang2023deepma, bao2023mdvsc, jia2023lightweight, lu_selfcritical_2024, liu2024cnn, liu2023transformer, zhou2021semantic,wang2022perceptual}. However, albeit the awfully exploded research interest, one critical question remains unsolved: \emph{why does the joint approach stand out, as separate source channel coding (SSCC) shall promise a greater degree of freedom from an optimization perspective?}  

    As the terminology implies, SSCC encompasses two decoupled ingredients: source coding and channel coding. The former part lies in effectively compressing the context, and the effectiveness of underlying Deep Neural Networks (DNN)-based predictors, such as Recurrent Neural Networks (RNN)-based DeepZip \cite{goyal2019deepzip}, Long Short Term Memory (LSTM)-based \cite{bellard2019lossless, liu2019decmac} and hybrid DNN-based Dzip \cite{goyal2021dzip}, have been validated widely in achieving satisfactory text compression. More prominently, Transformer-based \cite{vaswani2017attention} and Large Language Model (LLM)-based compression springs up recently \cite{mao2022fast, valmeekam2023llmzip, huang2023approximating, narashiman2024alphazip, mittu2024finezip}. The latest research \cite{deletang2024language} unveils the equivalence between compression and prediction. In other words, in the general framework where statistical models predict symbols and encoders use predictive probabilities to perform compression, better predictive models lead directly to better compressors \cite{deletang2024language}. Hence, the astonishing capability of LLM implies the potential for an unprecedented source codec.
    On the other hand, Error Correction Code (ECC) plays an indispensable role in channel coding. Although some advanced algebraic block codes like Bose–Chaudhuri–Hocquenghem (BCH) codes \cite{bose1960class}, Low-Density Parity-Check (LDPC) codes \cite{gallager1962low} and Polar codes \cite{arikan2009channel}, can somewhat ensure the reliability of transmission, the efficient decoding of ECC is an unresolved difficulty. Recently, DNNs have started to demonstrate their contribution in channel coding. For example, deep learning models are implemented to achieve Belief Propagation (BP) decoding \cite{nachmani2016learning, nachmani2018deep, nachmani2019hyper}, while a model-free Error Correction Code Transformer (ECCT) for algebraic block codes \cite{choukroun2022error} contributes to the enhancement of decoding reliability.

    In this paper, 
    on top of an LLM-based Arithmetic Coding (LLM-AC) system, the proposed SSCC framework integrates fine-grained, semantics-aware probability modeling and encoding  with ECCT-enhanced channel decoding, thus forming a closed-loop optimization framework. To the best of our knowledge, this work represents the very first comprehensive integration of LLM-based compression and ECCT-complemented channel decoding for a holistic SemCom architecture. Through extensively showcasing the performance superiority over JSCC, we argue 
    this performance improvement primarily arises after tackling the underlying incompatibility between conventional SemCom approaches \cite{xie2021deep, zhang2023deepma, bao2023mdvsc, jia2023lightweight, lu_selfcritical_2024, liu2024cnn, liu2023transformer, zhou2021semantic,wang2022perceptual} and digital communication architectures \cite{huang2024d}. Particularly, those approaches simply assume the deliverability of encoded semantic feature vectors while neglecting the energy costs associated with transmitting high-precision floating point numbers \cite{huang2024d}. However, further quantization \cite{tong_alternate_2023,tong_alternate_2024} and digital modulation can compromise the widely assumed existence of performance superiority in JSCC. Meanwhile, in contrast to the direct utilization of the astonishing semantic interpretation capability \cite{jiang2024semantic, liang2024generative, jiang2024large}, the deployment of LLMs focuses on the compression and encoding of text to squeeze the largely untapped redundancy. Therefore, our work is also significantly different from existing integrations of Generative AI (GAI) and SemCom \cite{grassucci2023generative, chang2024gensc, guo2023semantic, xie2024towards, qiao2024latency,jiang2024multi, yang2024rethinking}. Furthermore, the adoption of ECCT boosts the effectiveness of SSCC in specific cases. In a nutshell, our comprehensive evaluation of LLM and ECCT-based SSCC shows that \emph{separate source channel coding is still what we need}!

    The rest of this paper is organized as follows. Section \ref{sec:model} introduces the SSCC system model, while its key components are enumerated in Section \ref{sec:SSCC}. Numerical results are presented in Section \ref{sec:expeirments} to show the performance superiority of the proposed SSCC system. Finally, Section  \ref{sec:conclusion} concludes this paper with discussions on future works.
    For convenience, we list the major notations of this paper in Table \ref{Table:notations}.
    
    \begin{table}[t]
    \centering
    \caption{Major notations used in this paper.} \label{Table:notations}
    \begin{tabular*}{1.0\columnwidth}{c m{0.75\columnwidth}}
        \toprule[0.75pt]
        Notation    &    Definition    \\
        \midrule[0.5pt]
        $\bm{s}, \bm{\hat{s}}$    &    Source text sequence and its reconstructed version at the receiver side \\
        $\bm{t}, \bm{\hat{t}}$    &    Token sequence and its reconstructed version at the receiver side \\
        $C_s, C_e$    &    The source code and the channel code (error correction code) \\
        $\rho, \Tilde{\rho}$    &   The source distribution and the predicted probability distribution via LLM \\
        $\mathcal{D}, D_i, \tau$    &    Source code dictionary, its $i$-th character and total vocabulary size \\
        $\mathbb{I}_k, l_k, u_k$    &    Probability interval at step $k$ of source coding and its corresponding lower bound and upper bound \\
        $\bm{m}, \bm{\hat{m}}$    &   Source-coded message vector and its channel-decoded version \\
        $\lambda$    &    Probability interval represented in decimal form \\
        $N, K$    &    Codeword length and message length of error correction code $C_e(N,K)$ \\
        $\bm{G,H}$    &    Generator matrix and parity check matrix \\
        $\bm{x}, \bm{x}_b, \bm{x}_s$    &    Transmitted codeword vector encoded by the channel coder, its binary representation (taking values in {0,1})and its signed representation (taking values in {-1,+1}) \\
        $\bm{\hat{x}}, \bm{\hat{x}}_b$    &    Soft approximation of codeword, and its binary representation \\
        $\mathcal{N}(\cdot,\cdot), \sigma_n$    &    Gaussian distribution and noise standard deviation \\ 
        $h$    &    Channel fading coefficient \\
        $\bm{z}, \bm{\tilde{z}}, \bm{\hat{z}}$    &    Additive Gaussian noise, corresponding multiplicative noise and ECCT prediction result  \\
        $\bm{y}, \bm{y}_b, \bm{{\Tilde{y}}}$    &    Received noisy codeword, its binary representation and pre-processed version \\
        $\syn(\cdot)$    &    Syndrome computation function defined in ECCT \\
        $f(\cdot)$    &    ECCT decoding function \\
        $\bm{W}$    &    Learnable embedding matrix for high-dimensional mapping \\
        $\bm{g}(\cdot)$    &    Code-aware self-attention mask function \\
        \bottomrule[0.75pt]
    \end{tabular*}
    \end{table}



\section{System Model}
\label{sec:model}
    Our SSCC framework encompasses the following ingredients.
    \begin{itemize}
        \item \emph{Source Encoding}: The input text sequence denoted as $\bm{s}_{1:N_s}$ undergoes a source encoder that converts characters into a compressed binary message $\bm{m} \in \{0,1\}^K$. During source encoding, Arithmetic Coding (AC) can be leveraged for effective compression here. For LLM-based processing, an intermediate result (i.e., a sequence of tokens $\bm{t}_{1:N_t}$) can be obtained during the transformation from $\bm{s}$ to $\bm{m}$. 
        \item \emph{Channel Encoding and Modulation}: The message $\bm{m}$ is then encoded via a LDPC code $C_e(N,K)$, which is selected for its excellent error-correction capabilities and compatibility with iterative decoding algorithms, as mentioned in \cite{choukroun2022error}. The encoding process employs a generator matrix $\bm{G}$ to transform the message  in $\bm{m}$ to a codeword $\bm{x} \in \{0,1\}^N$. The parity check matrix $\bm{H}$, which satisfies $\bm{G}\cdot\bm{H}^T=0$ and $\bm{H}\cdot\bm{x}=0$ is a key component of the LDPC decoding process. Afterwards, Binary Phase Shift Keying (BPSK) modulation maps the binary codeword $\bm{x}$ to a sequence of symbols $\bm{x}_s \in \{0,1\}^N$ suitable for transmission over the wireless channel. Notably, other error correction codes such as Polar codes \cite{arikan2009channel} can be applied as well.
        \item \emph{Channel}: The modulated signal $\bm{x}_s$ is transmitted over a noisy channel, modeled as an Additive White Gaussian Noise (AWGN) channel or a Rayleigh fading channel. The received signal $\bm{y} \in \mathbb{R}^N$ is corrupted by additive noise $\bm{z} \sim \mathcal{N}(0,\sigma_n^2)$, resulting in $\bm{y}=h\bm{x_s}+\bm{z}$, where $h$ is the channel fading coefficient.
        \item \emph{Demodulation and Channel Decoding}: BPSK demodulation recovers a binary codeword $\bm{\hat{x}}_b\in\{0,1\}^N$ from $\bm{\hat{x}}$. Subsequently, channel decoder reconstructs the message $\bm{\hat{m}} \in \{0,1\}^K$ from $\bm{\hat{x}}_b$. In contrast to conventional approaches that employ either hard-decision (e.g., bit-flipping algorithm) or soft-decision (e.g., sum-product algorithm) algorithms to decode LDPC codewords transmitted through the channel, some complementary decoding modules, such as ECCT, can be applied prior to demodulation to enhance the decoding performance. Notably, ECCT can provide an estimation of the transmitted codeword $\bm{\hat{x}}_b$, denoted as $\bm{\hat{x}}$, while subsequent demodulation and information bits extraction are then performed on the estimated codeword $\bm{\hat{x}}$.
        \item \emph{Source Decoding}: The recovered message $\bm{\hat{m}}$ is ultimately decoded by the source decoder, which reconstructs the original text sequence $\bm{s}$ from the message, effectively reversing the encoding process. Similar to the encoder, the decoder can implement arithmetic decoding.
    \end{itemize}
    In comparison, JSCC often leverages an end-to-end DNN to implement source and channel codecs. Here, the terminology ``end-to-end'' implies the generally joint training of source and channel codes, as adopted in most works. The further details on JSCC can be found in \cite{lu_semanticsempowered_2024} and the references therein. In the following section, we will address how to leverage the strength of LLM to enhance text compression and reconstruction, combined with the robustness of ECCT-complemented LDPC codes for error correction, as shown in Fig. \ref{fig:System_Model}.


\section{Proposed SSCC Framework}
\label{sec:SSCC}
In this section, we introduce LLM-based source coding and ECCT-complemented channel coding.
\subsection{LLM-based Source Coding}
    \begin{figure*}
        \centering
        \includegraphics[width=.75\textwidth]{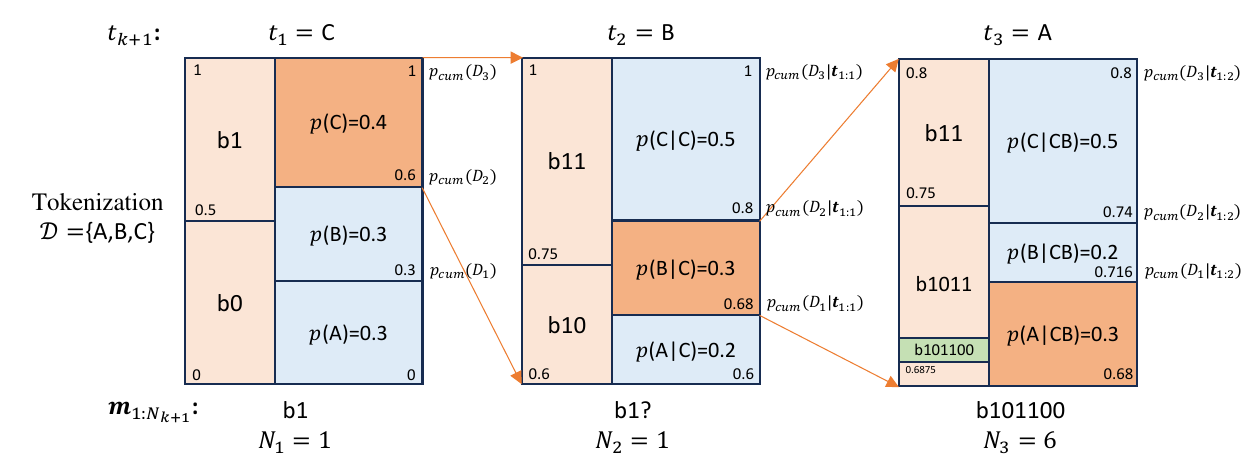}
        \caption{An example of arithmetic coding.}
        \vspace{-0.5cm}
        \label{fig:AC example}
    \end{figure*}
    
    Given a source distribution $\rho$, lossless compression aims to encode a text sequence $\bm{s}$ sampled from $\rho$ into a binary code $\bm{m} = C_s(\bm{s})$ of minimal possible length with no loss of original information. According to Shannon's source coding theorem \cite{shannon1948mathematical}, the optimal expected bit length is $L_{\min}=\mathbb{E}_{\bm{s}\sim\rho}[-\log_2\rho(\bm{s})]$. To obtain such optimal length, arithmetic coding \cite{rissanen1976generalized, pasco2006source}, a form of entropy encoding, is typically adopted contingent on a probabilistic model over $\rho$ or its marginal distribution.
    Arithmetic coding implies that frequently used characters are stored with fewer bits while rarely occurred characters correspond to more bits, resulting in fewer bits used in total.

    In particular, the input text sequence denoted as $\bm{s}$ undergoes tokenization by the LLM tokenizer, which converts characters into a sequence of tokens $\bm{t}$ for LLM to process. The LLM subsequently generates a compact representation of the text, effectively encoding the tokens into a compressed binary message $\bm{m} \in \{0,1\}^K$. 
    Specially, considering a dictionary $\mathcal{D}$ of $\tau$ tokens, the input sequence $\bm{s}$ is first parsed into token sequence $\bm{t}$. 
    Given the first $k$ tokens $\bm{t}_{1:k}$, the $(k+1)$-th token $t_{k+1}$ can be inferred as a predicted probability distribution $\tilde{\rho}(t_{k+1}|\bm{t}_{1:k})$. Here, $\tilde{\rho}(t_{k+1}|\bm{t}_{1:k})$ indicates the LLM's estimation of the true distribution $\rho(t_{k+1}|\bm{t}_{1:k})$.
    The incremental decoding nature in LLM implies that it can accurately predict the probability distribution of the next token based on known ones, thereby providing a sub-optimal estimation of the true distribution \cite{deletang2024language}. As shown in Fig. \ref{fig:AC example}, selecting the next character is actually narrowing down the probabilistic interval where the sequence is located, which means the code $\bm{m}$ is determined once the interval is fixed.
    Starting with $\mathbb{I}_0 = [0,1)$, the previous interval determined by $\bm{t}_{1:k}$ in step $k$ is defined as $\mathbb{I}_k = [l_k, u_k)$. 
    Therefore, denoting $p(t_{k+1}=D_j) =  \tilde{\rho}(t_{k+1} = D_j|\bm{t}_{1:k})$, 
    
    \begin{equation}
    \begin{aligned}
        \mathbb{I}_{k+1}(D_i) = \Big[ &l_k + (u_k - l_k) \times \sum\nolimits_{j < i}p(t_{k+1}=D_j), \\
        &l_k + (u_k - l_k) \times \sum\nolimits_{j \leq i}p(t_{k+1}=D_j) \Big).\\
    \end{aligned}
    \end{equation}
    
    In practice, we consider finite precision arithmetic encoders referring to \cite{howard1994arithmetic}, with pseudo-code provided in Appendix \ref{app:code}. 
    Consequently, we can obtain a binary code $\bm{m}=C_s(\bm{s}_{1:N_s})$ of the shortest length,  completely corresponding to the probability interval determined by the sequence. 
    At the receiver side, if the receiver shares a consistent source distribution $\tilde{\rho}$ with the sender, given the received (and channel decoded) bit sequence $\hat{\bm{m}}$ corresponding to $C_s(\bm{s}_{1:N_s})$, 
    we can decode $t_{K_{n+1}}=D_i \in \mathcal{D}$ when the first n bits have been decoded by identifying $D_i$ such that 
    \begin{equation}
    \begin{aligned}
        & \hat{\mathbb{I}}_{n+1} = [l_{n+1},u_{n+1}) = 
        \begin{cases}
            \left[ l_n, \frac{1}{2}(l_n+u_n) \right), \text{if } m_{n+1}=0\\
            \left[ \frac{1}{2}(l_n+u_n), u_n \right), \text{if } m_{n+1}=1\\
        \end{cases} \\
        \subseteq 
        & \hat{\mathbb{I}}_{K_{n+1}}(D_i) 
        \!\begin{aligned}[t]
            &= [L,U) \\
            &= 
            \!\begin{aligned}[t]
                \Big[ & l_{K_{n+1}} + (u_{K_{n+1}} - l_{K_{n+1}}) \times \sum\nolimits_{j < i}p(D_j), \\
                & l_{K_{n+1}} + (u_{K_{n+1}} - l_{K_{n+1}}) \times \sum\nolimits_{j \leq i}p(D_j) \Big).\\
            \end{aligned}
        \end{aligned}
    \end{aligned}
    \end{equation}
    where $p(D_j)$ represents $p(t_{K_{n+1}+1}=D_j)$. For more details, please refer to Appendix \ref{app:code}. 

    \begin{figure*}
        \centering
        \includegraphics[width=.85\textwidth]{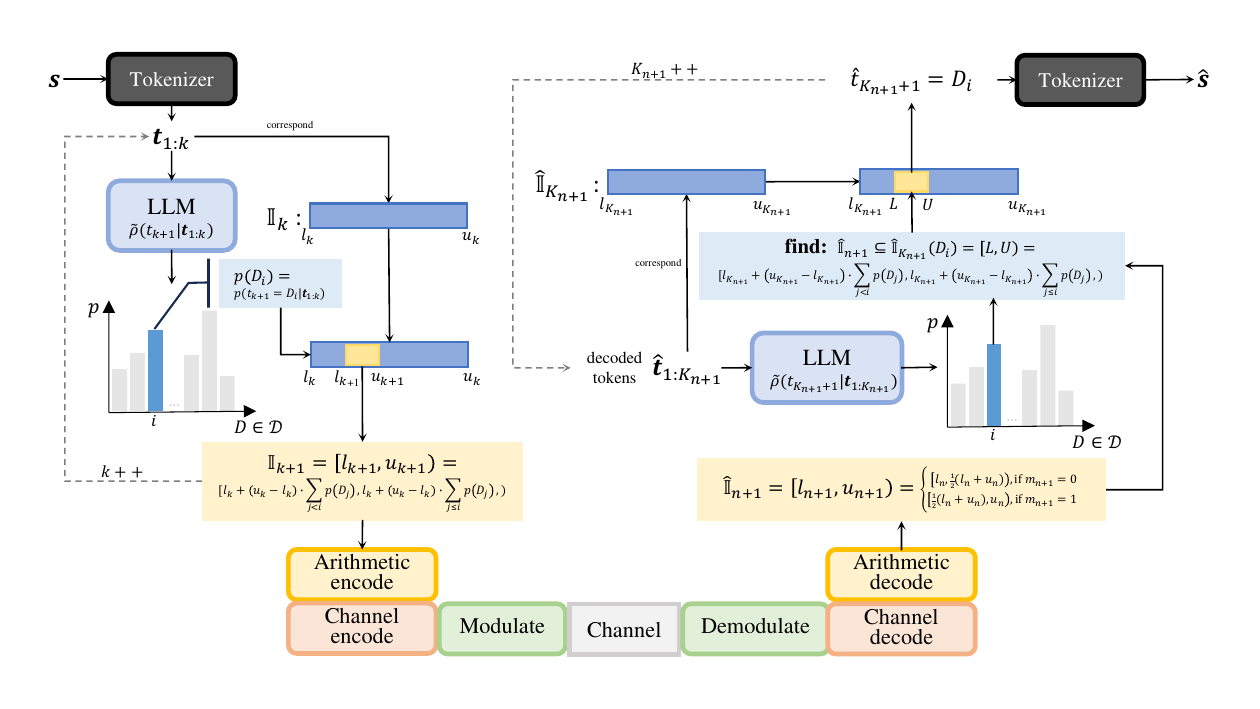}
        \caption{LLM-based arithmetic encoding and decoding.}
        \label{fig:LLM-based_AC_encoder}
    \end{figure*}
    Fig. \ref{fig:LLM-based_AC_encoder} illustrates such LLM-based arithmetic encoding and decoding where the LLM provides a probability interval according to the text sequence $\bm{s}$.  Unlike the online setting, which trains the model on the data to be compressed, this paper assumes the availability of a well-trained LLM and employs it to compress different datasets, following the offline setting used in \cite{valmeekam2023llmzip}.
    

    \begin{remark}
        \cite{deletang2024language} figures out that the expected code length achieved by leveraging LLM as a compressor could be represented as the cross-entropy, that is,
    \begin{equation}
        H(\rho, \tilde{\rho}) := \mathbb{E}_{\bm{s}\sim\rho}\left[\sum\nolimits_{i=1}^{n}-\log_2\tilde{\rho}(s_i|\bm{s}_{<i})\right]
        \label{eq:ce_compressor},
    \end{equation}
    where $\rho$ is the source distribution and $\tilde{\rho}$ is the estimation of $\rho$ via a parametric probabilistic model. Hence, the compression shares the same training objective as prediction. Therefore, it can be interpreted as the link between the model log-loss and the compression rate, providing theoretical support for the employment of LLM for source coding. 
    \end{remark}

\subsection{Error Correction Code Transformer}

    ECCT \cite{bennatan2018deep} belongs to a complementary transformer-alike module to ensure the channel decoding reliability. Notably, ECCT involves 
    specific preprocessing and post-processing steps to effectively avoid overfitting. Without loss of generality, before preprocessing, the syndrome of codes is defined by
    \begin{equation}
    \begin{aligned}
        \syn(\bm{y}) := & \bm{H y_b} = \bm{H}\signbin(\bm{y}) \\
        = &\frac{1}{2}\bm{H}(1-\sign(\bm{y})) \in \{ 0,1 \}^{N-K}.\\
    \end{aligned}
    \end{equation}
    This should be checked first upon receiving the signal, since corruption could be detected immediately if $\syn(\bm{y})$ is a non-zero vector. In other words, an all-zero syndrome ensures that the received signal suffers no distortion. Note that the function $\signbin(\text{·})$ could be viewed as a hard decision on $\bm{y}$, and $\sign(\text{·})$ here denotes a sign function defined by
    \begin{equation}
        \sign(y) = 
        \begin{cases}
                1, & \quad y > 0; \\
                0, & \quad y = 0; \\
                -1, & \quad y < 0.
        \end{cases}
    \end{equation}
    Next, ECCT constructs a $2N-K$ dimensional input embedding by concatenating the element-wise magnitude and syndrome vectors, such that
    \begin{equation}
        \Tilde{\bm{y}}:=\big[ |\bm{y}|, \syn(\bm{y}) \big] \in \mathbb{R}^{2N-K},
    \end{equation}
    where $[\text{·},\text{·}]$ denotes vector/matrix concatenation and $|\bm{y}|$ denotes the absolute value (magnitude) of $\bm{y}$. 
    
\begin{figure*}
        \centering
        \includegraphics[width=.9\textwidth]{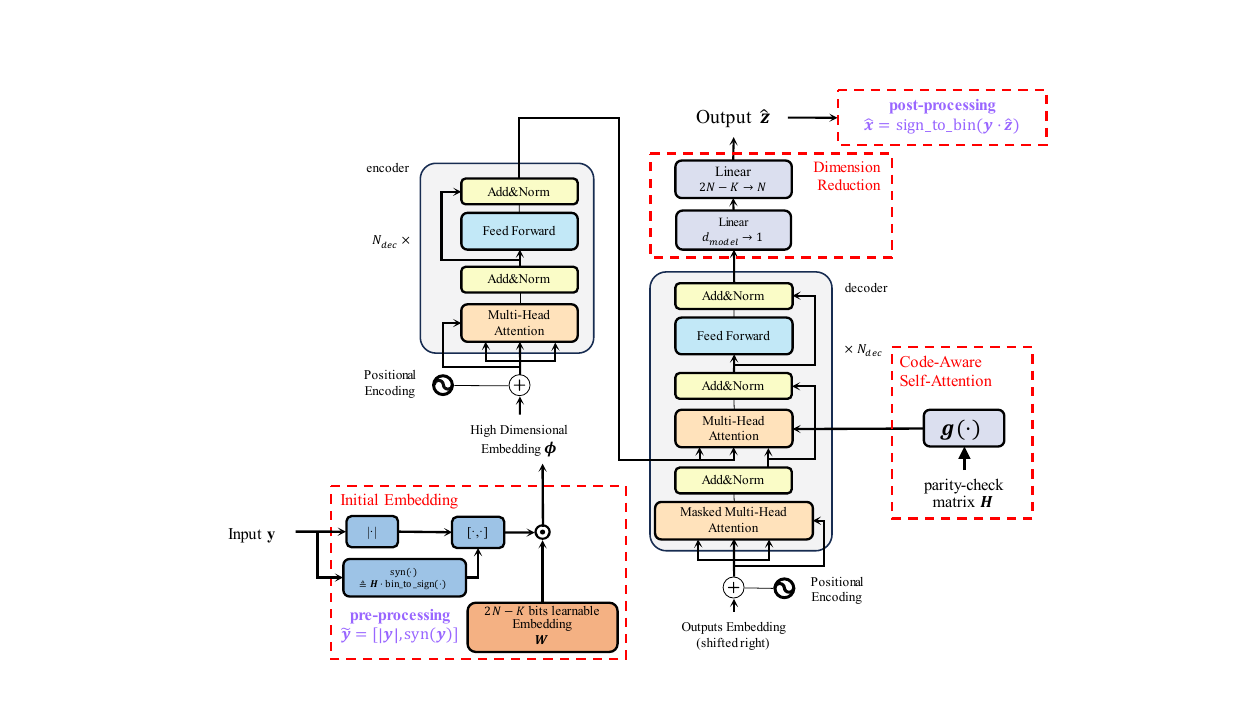}
        \caption{ECCT architecture.}
        \label{fig:ECCT}
    \end{figure*}

    The objective of the decoder is to predict the multiplicative noise $\Tilde{\bm{z}}$ from $\bm{y}$, where $ \bm{y} = h\bm{x}_s+\bm{z} = \bm{x}_s (h + \bm{x}_s\bm{z}) = \bm{x}_s \bm{\Tilde{z}} $.
    Compared to traditional Transformer architectures \cite{vaswani2017attention}, ECCT introduces two additional modules for positional reliability encoding and code aware self-attention, as shown in Fig. \ref{fig:ECCT}. Note that ECCT processes the channel output $\bm{y}$ as input and generates a prediction $\hat{\bm{z}}$ of the multiplicative noise $\Tilde{\bm{z}}$. The key differences between ECCT and traditional Transformer architectures are highlighted in the red box in Fig. \ref{fig:ECCT}. Implementations details are provdided in Appendix \ref{app:ecct}.

    Finally, the training process aims to minimize the Binary Cross Entropy (BCE) loss between the predicted noise $\hat{\bm{z}}$ and the multiplicative noise $\Tilde{\bm{z}}$, given by
    \begin{equation}
    \begin{aligned}
        \text{loss} &= \texttt{BCELoss}(\hat{\bm{z}}, \Tilde{\bm{z}}) \\
        &= -\frac{1}{N} \sum\nolimits_i\Big(\texttt{bin}(\Tilde{\bm{z}}_{i}) \cdot  \log\big(\sigma(\hat{\bm{z}}_{i})\big) \\
        &\quad + \big(1-\texttt{bin}(\Tilde{\bm{z}}_{i})\big) \cdot \log\big(1-\sigma(\hat{\bm{z}}_{i})\big)\Big),\\
    \end{aligned}
    \end{equation}
    where $\sigma(\cdot)$ denotes the sigmoid activation function.
    \begin{remark}
    The estimation of multiplicative noise is represented as $\hat{\bm{z}} = f(\Tilde{\bm{y}})$, while the post-processing step estimates $\bm{x}$ by $\hat{\bm{x}} = \signbin(\bm{y} \cdot f(\Tilde{\bm{y}}))$. Given that, for correct estimation, $\sign(\hat{\bm{z}}) = \sign(\Tilde{\bm{z}})$. Therefore, 
    \begin{equation}
    \begin{aligned}
        \hat{\bm{x}} &= \signbin(\bm{y} \cdot f(\Tilde{\bm{y}})) \\
        &= \signbin(\bm{x_s\Tilde{\bm{z}}\cdot\hat{\bm{z}}}) \\
        &= \signbin(\bm{x_s}) = \bm{x}.\\
    \end{aligned}
    \end{equation}
    In other words, ECCT contributes to noise-free channel coding.
    \end{remark}


\section{Experiments}
\label{sec:expeirments}

    In this section, we compare the proposed method with traditional SSCC approaches and existing JSCC solutions under both AWGN and Rayleigh fading channels.

    \begin{table}[t]
    \centering
    \begin{threeparttable}
    \caption{Mainly used hyperparameters in the experiments.} \label{Table:setting}
    \begin{tabular*}{.8\columnwidth}{ccc}
        \toprule[0.75pt]
        Model                     &   Hyperparameter   &  Value    \\
        \midrule[0.5pt]
        \multirow{5}{*}{ECCT}     &   Learning rate    &   $10^{-4}$   \\
    	                       &    Batch size    &   $128$   \\
    	                       &    Number of decoder layers   &   $6$   \\
    	                       &    Dimension of embedding  &   $32$   \\
    	                         &    Number of attention heads  &   $8$   \\
        \midrule[0.5pt]
        \multirow{6}{*}{DeepSC}     &    Learning rate   &   $10^{-4}$   \\
    	                          &    Batch size    &   $64$   \\
    	                        &    Number of encoder/decoder layers    &   $4$   \\
    	                        &    Dimension of embedding  &   $128$   \\
                                    &    Dimension of FFN\tnote{1}    &   $512$   \\
    	                          &    Number of attention heads  &   $8 $  \\
        \midrule[0.5pt]
        \multirow{6}{*}{UT}     &    Learning rate   &   $10^{-4}$   \\
    	                      &    Batch size    &   $64$   \\
    	                    &    Number of encoder/decoder layers  &   $3$   \\
    	                    &    Dimension of embedding   &   $128$   \\
                                &    Dimension of FFN\tnote{1}  &   $1,024$   \\
    	                      &    Number of attention heads  &   $8$   \\
        \bottomrule[0.75pt]
    \end{tabular*}
    \begin{tablenotes}
        \item[1] FFN represents Feed Forward Network
    \end{tablenotes}
    \end{threeparttable}
    \end{table}

\subsection{Simulation Settings}
    To facilitate comparison, we utilize a pre-processed dataset consisting of the standard proceedings of the European Parliament \cite{koehn2005europarl}. A segment of this dataset is selected as an example and fed as the source to a Generative Pre-trained Transformer (GPT)-2 \cite{radford2019language} model for source coding. In this numerical experiment, we primarily choose the smallest GPT2-base model with $124$ million parameters, while larger models (e.g., the $355$-million-parameter GPT2-medium, the $774$-million-parameter GPT2-large, and the $1.5$-billion-parameter GPT2-XL) are subsequently used for comparative analysis. Arithmetic coding based on the LLM is configured with a precision limit of $31$ bits. For channel coding, we adopt an LDPC code with an information word length of $24$ and a codeword length of $49$, denoted as $\text{LDPC}(49,24)$, resulting in a code rate close to $1/2$. Subsequently, ECCT is used for algebraic block code decoding, which is capable of training on diverse error correction codes. The hyperparameter settings for ECCT training are detailed in Table \ref{Table:setting}. For comparative analysis, we select DeepSC \cite{xie2021deep}, UT \cite{zhou2021semantic}, UT with quantization\footnote{Compared to DeepSC \cite{xie2021deep} and UT \cite{zhou2021semantic}, which directly transmit the encodes floats, UT with quantization maps the encoding results to a fixed number ($30$) of bits for transmission.} as benchmark JSCC algorithms. Considering the subsequent Signal-to-Noise Ratio (SNR) performance comparison, both algorithms are trained using mixed precision (i.e., \texttt{float16}), which, as discussed later, has a minimal negative impact on SNR computation. Key parameters used for training DeepSC and UT are also listed in Table \ref{Table:setting}. Besides, the traditional approach employs Huffman coding for source coding. Furthermore, Bilingual Evaluation Understudy (BLEU) \cite{papineni2002bleu}, and semantic similarity measured by BERT \cite{kenton2019bert} are used to measure performance, as these metrics are widely recognized in natural language processing.

    Most existing SemCom works evaluate the performance with respect to the $\text{SNR} = 10\log_{10}(\frac{E_{\text{tb}}}{N_0})$/dB, where $E_{\text{tb}}$ denotes the energy associated with transmitting a single bit after source/channel coding and digital modulation, and $N_0$ represents the noise power spectral density. However, since different coding and modulation schemes across different communication methodologies result in varying numbers of bits transmitted over the physical channel, such a comparative metric of $\text{SNR}$ ignores the differences in delivering different numbers of bits. Instead, referring to the total energy consumption $E_{\text{total}}$ by sending $\text{Num}_\text{unified}$ bits through the physical channel in an LLM-based SSCC system, we propose a consistent definition of SNR, in terms of an LLM-based SSCC reference baseline $\text{SNR}_{\text{unified}}$, as a function of the practical employed bits $\text{Num}$. Mathematically, this is expressed as:
    \begin{align}
        \text{SNR} 
        &= 10\log_{10}(\frac{E_{\text{total}}}{N_0 \cdot \text{Num}}) \nonumber\\
        &= 10\log_{10}(\frac{E_{\text{total}}}{N_0 \cdot \text{Num}_\text{unified}} \times \frac{\text{Num}_\text{unified}}{\text{Num}}) \\
        &= \text{SNR}_{\text{unified}} + 10\log_{10}(\frac{\text{Num}_\text{unified}}{\text{Num}}).\nonumber
    \end{align}
    where $\text{SNR}_{\text{unified}}$ is used as an independent variable for aligning $E_{\text{total}}$, while for bit-oriented transmission (resp. float-based JSCC), $\text{Num}$ denotes the number of bits (resp. float vectors) transmitted through the channel.
    
    On the other hand, as mentioned in Section \ref{sec:intro} and \cite{huang2024d}, deep learning-based JSCC systems extract the semantic feature of information to embed vectors in latent space, which is incompatible with digital communication systems. For JSCC methodologies like UT \cite{zhou2021semantic} and DeepSC \cite{xie2021deep}, transmitting a float number certainly consumes far more energy than delivering a binary bit. In this case, if \texttt{float16} is adopted, we can roughly assume it consumes an additional $10\times \log_{10} (16) \approx 12.041$/dB. Hence, for the float-based JSCC methods, the \emph{unified} evaluation metric is further modified to maintain a consistent energy consumption across different methodologies. 
    In summary,
    \begin{align}
        &\text{SNR} \\
        & = \begin{cases}
            \text{SNR}_{\text{unified}} + 10\log_{10}(\frac{\text{Num}_\text{unified}}{\text{Num}}) + 12.041, & \text{float-based};\\
            \text{SNR}_{\text{unified}} + 10\log_{10}(\frac{\text{Num}_\text{unified}}{\text{Num}}), & \text{otherwise}.
        \end{cases} \nonumber
    \end{align}
    During evaluation, experiments are conducted for different schemes in terms of $\text{SNR}_{\text{unified}}$.

\subsection{Numerical Results}
\begin{figure*}[tb]
        \centering
        \includegraphics[width=\textwidth]{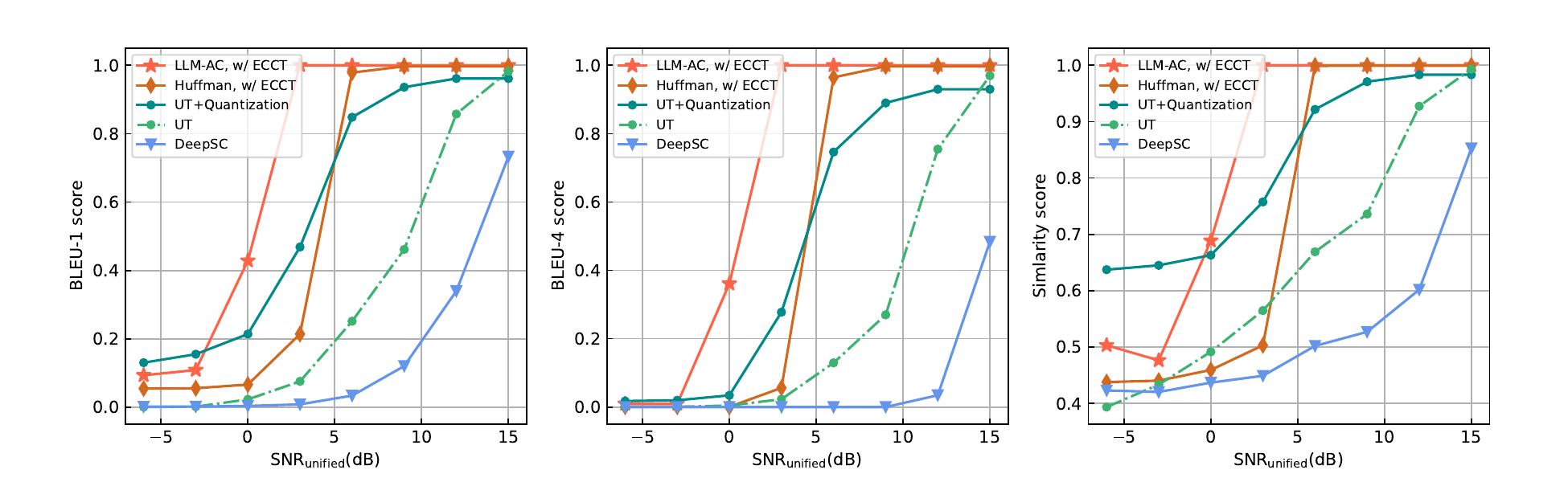}
        \caption{BLEU and Similarity scores versus $\textbf{SNR}_{\textbf{unified}}$ are evaluated for the same number of transmitted symbols. The proposed LLM-based SSCC is compared with Huffman coding with $\text{LDPC}(49,24)$ in BPSK, DeepSC, UT, and UT with quantization under the AWGN channel.}
        \label{fig:Comparison_AWGN}
    \end{figure*}

    In this section, 
    we implement the GPT2-base model as compressor and ECCT-complemented $\text{LDPC}(49,24)$ as the error correction code, and  
    compare it with DeepSC \cite{xie2021deep}, UT \cite{zhou2021semantic}, UT with quantization \cite{zhou_adaptive_2022} and the classical SSCC encompassing Huffman coding and ECCT (Fig. \ref{fig:Comparison_AWGN}). The results clearly demonstrate the superior performance of the proposed SSCC over the other three schemes.
    \begin{figure*}[tb]
        \centering
        \includegraphics[width=\textwidth]{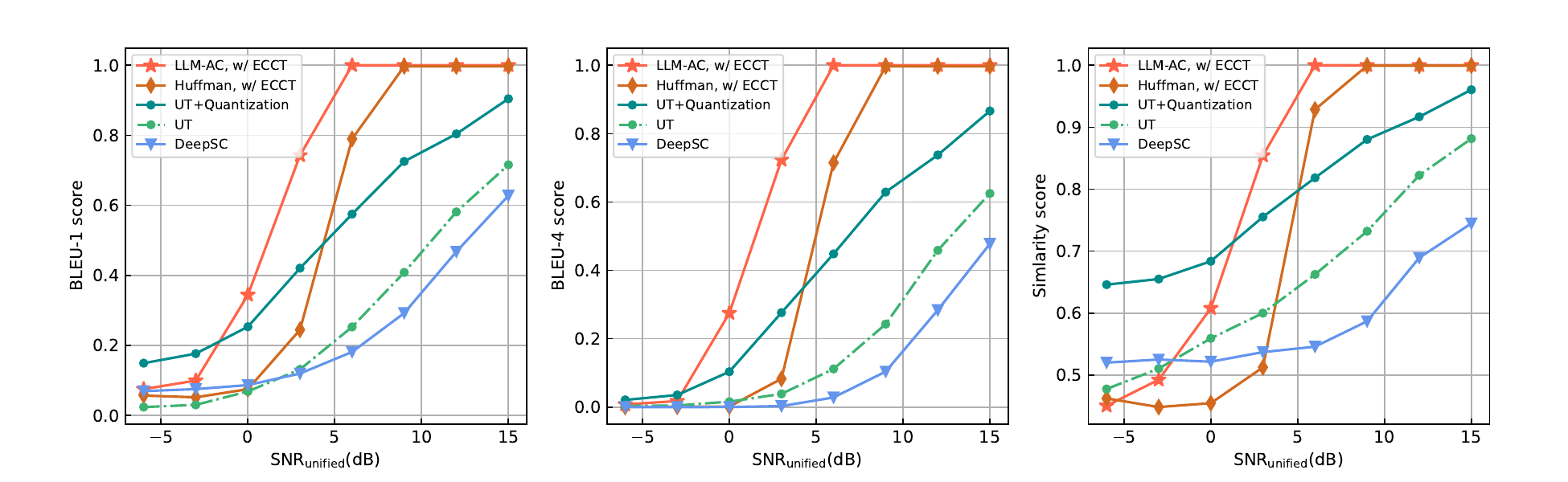}
        \caption{BLEU and Similarity scores versus $\textbf{SNR}_{\textbf{unified}}$ for the same number of transmitted symbols. The proposed LLM-based SSCC is compared with Huffman coding with $\text{LDPC}(49,24)$ in BPSK, DeepSC, UT, and UT with quantization under the Rayleigh fading channel.}
        \label{fig:Comparison_Rayleigh}
    \end{figure*}
    Similarly, we evaluate the performance under a Rayleigh fading channel in 
    Fig. \ref{fig:Comparison_Rayleigh}, where the results show that our system has a clear advantage in terms of the word-level BLEU score. However, in terms of semantic similarity, both the LLM-based and the traditional Huffman-based SSCC systems exhibit some disadvantages at lower SNRs, but still maintain a noticeable advantage at high SNRs.
    
    \begin{figure*}[tb]
        \centering
        \includegraphics[width=\textwidth]{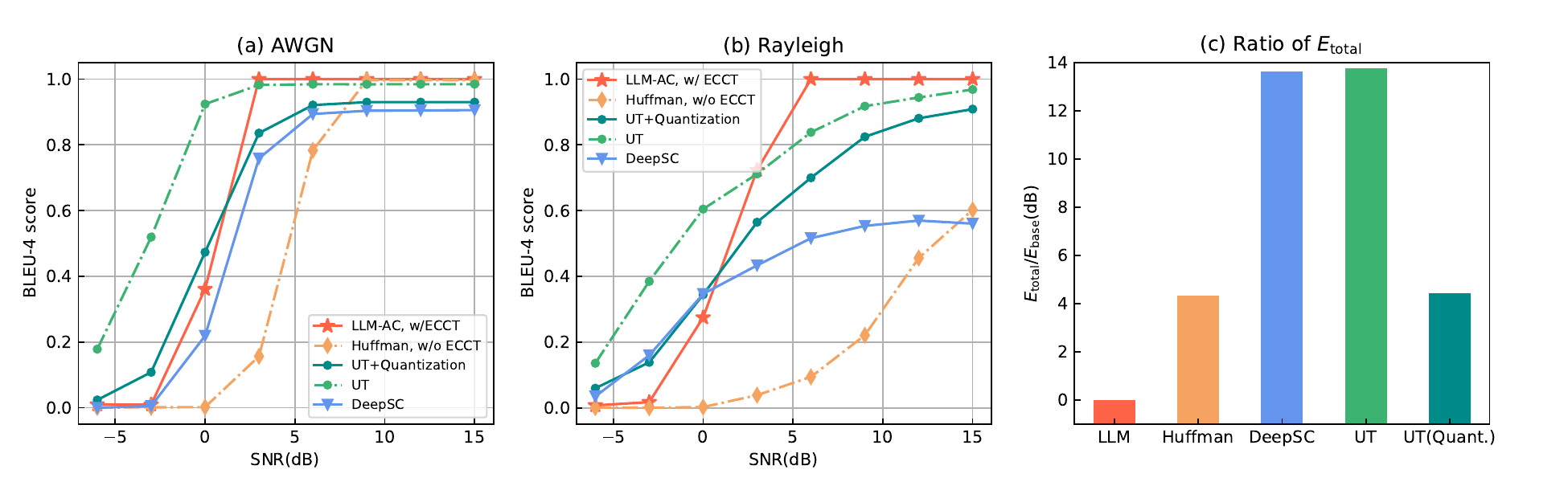}
        \caption{BLEU-4 score versus $\textbf{SNR}$ is evaluated for the same number of transmitted symbols. The proposed LLM-based SSCC is compared with Huffman coding with $\text{LDPC}(49,24)$ in BPSK (without ECCT); DeepSC, UT, and UT with quantization under (a) AWGN and (b) Rayleigh fading channels; (c) shows the ratio of $E_{\text{total}}$ among different systems.}
        \label{fig:Comparison_ununified_SNR}
    \end{figure*}
    
    In addition to presenting our key experimental results with $\text{SNR}_{\text{unified}}$ as the alignment metric, Fig. \ref{fig:Comparison_ununified_SNR} provides performance comparisons using traditional $\text{SNR}$ alignment, as well as the ratio of $E_{\text{total}}$ used by different systems over our LLM-based solution. This illustrates the additional energy consumption of JSCC systems in achieving superior performance.
    Apparently, JSCC systems in SemCom achieve significant gains mainly due to the extra energy consumption.
    
    \begin{figure*}[tb]
        \centering
        \includegraphics[width=.75\textwidth]{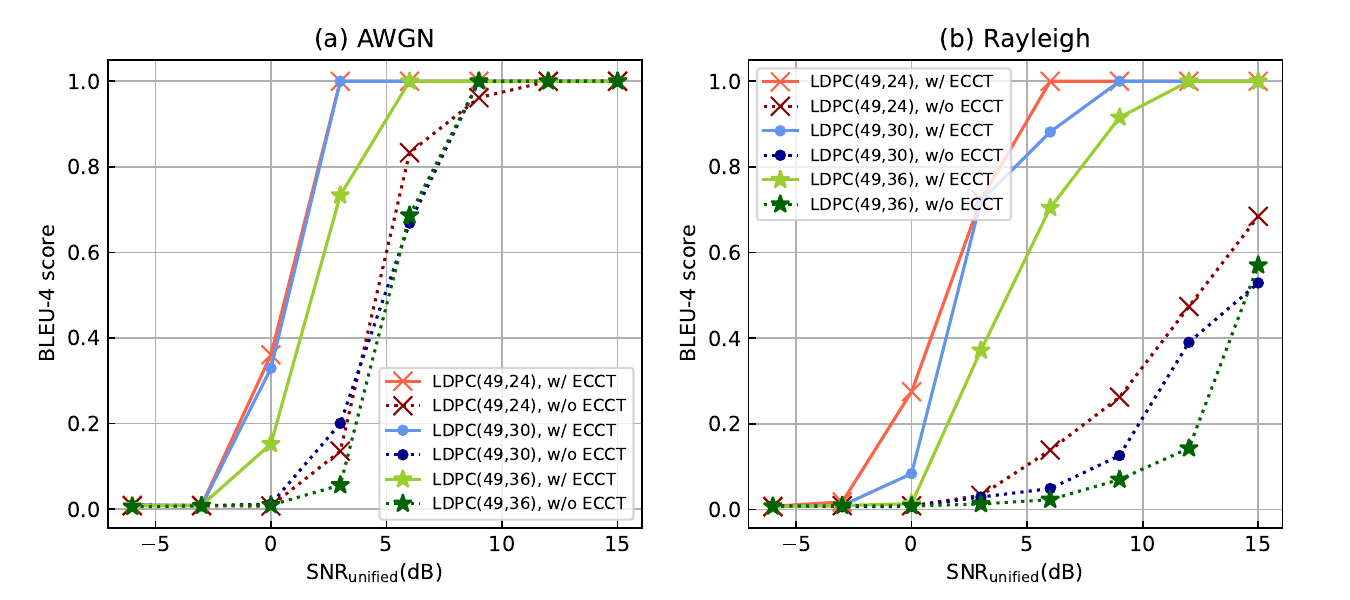}
        \caption{BLEU-4 score versus $\textbf{SNR}_{\textbf{unified}}$ for the same number of transmitted symbols, with different code rates using $\text{LDPC}(49,24)$ / $\text{LDPC}(49,30)$ / $\text{LDPC}(49,36)$ in BPSK, compared with the situations removing ECCT, under (a) AWGN and (b) Rayleigh fading channels.}
        \label{fig:LDPCrate_ECCTablation}
    \end{figure*}

    Afterward, we validate the contributing effectiveness of ECCT \cite{choukroun2022error} by
    comparing the performance of error correction codes with different coding rates under the same code length. Without loss of generality, the evaluation results based on LDPC under AWGN and Rayleigh fading channel are given in Fig. \ref{fig:LDPCrate_ECCTablation}. Notably, while the work in \cite{choukroun2022error} does not include Rayleigh channel results, inspired by the subsequent work on Denoising Diffusion Error Correction Codes (DDECC, \cite{choukroun2022denoising}), we extend ECCT to Rayleigh channels in a similar manner.
    It can be observed from Fig. \ref{fig:LDPCrate_ECCTablation} that compared to traditional LDPC decoding methods such as bit-flipping, ECCT provides consistent performance improvements.
    Furthermore, for error correction codes of the same length, lower coding rates demonstrate better recovery of noisy signals under the same SNR. More importantly, without ECCT, traditional algorithms struggle to decode noisy signals under Rayleigh channels effectively, and reducing the coding rate slightly improves the performance trivially. However, ECCT trained under the Rayleigh channel achieves as competitive performance as that under the AWGN channel.
    
\begin{figure*}[tb]
        \centering
        \includegraphics[width=0.875\textwidth]{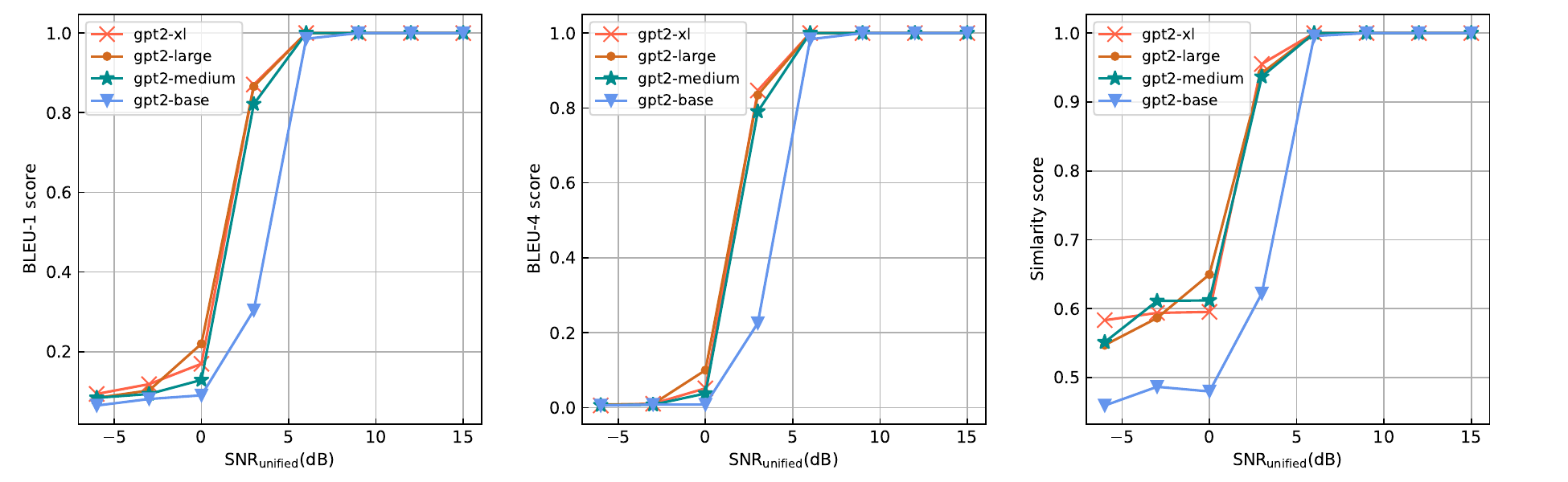}
        \caption{BLEU and Similarity scores of models with versus $\textbf{SNR}_{\textbf{unified}}$, with different parameter scales (GPT2, GPT2-medium, GPT2-large, GPT2-XL), using $\text{LDPC}(121,110)$ as the error correction code.}
        \label{fig:scaling}
    \end{figure*}

\begin{figure}[tb]
        \centering
        \includegraphics[width=0.75\columnwidth]{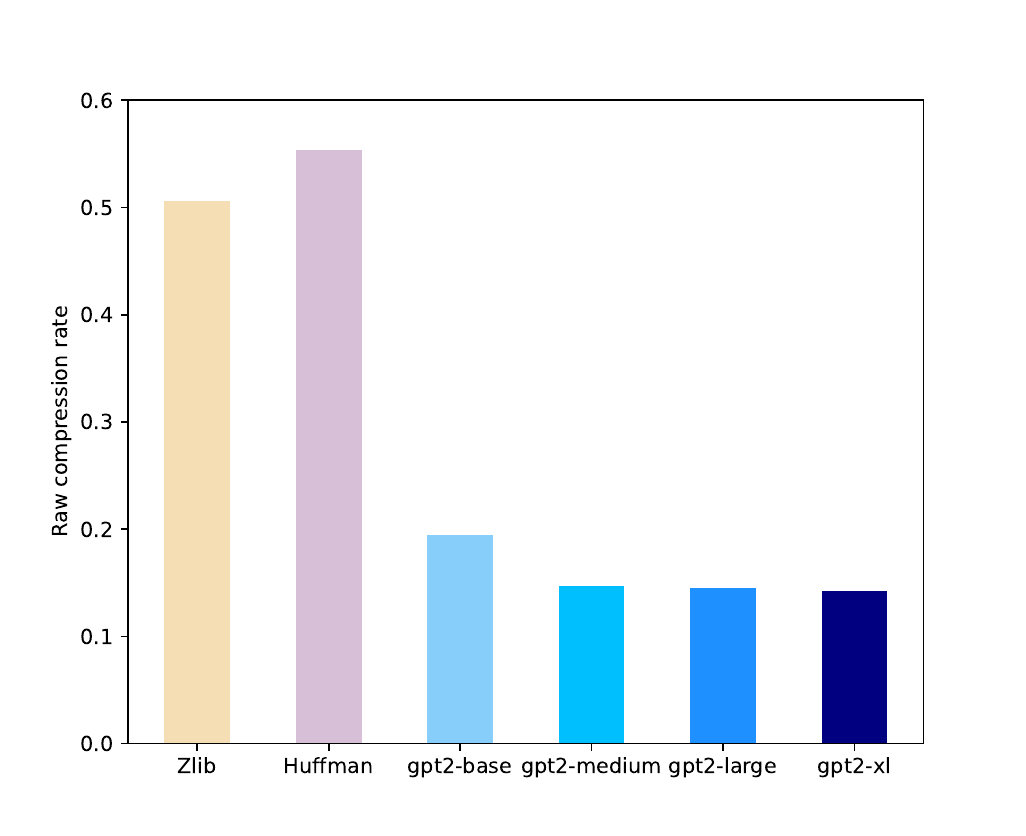}
        \caption{Compression rate comparison between traditional methods (Zlib and Huffman coding) and LLM-AC.}
        \label{fig:compress_rate}
    \end{figure}

    Considering the scaling law and emergent abilities of LLMs, we evaluate the performance by combining different models from the GPT-2 family with ECCT-complemented LDPC channel coding (i.e., a high-rate $\text{LDPC}(121,110)$ code). Both the end-to-end SSCC performance in Fig. \ref{fig:scaling} and the compression rate in Fig. \ref{fig:compress_rate} indicate a notable performance improvement after adopting a model larger than GPT2. However, the performance difference among GPT2-medium, GPT2-large, and GPT2-XL is marginal. We hypothesize that while increasing the model size beyond a certain threshold contributes significantly to system performance, variations within a specific range of model scales yield diminishing returns.
    Furthermore, inspired by \cite{huang2024compression}, 
    the performance comparison with Zlib and static Huffman coding in Fig. \ref{fig:compress_rate} 
    demonstrates that LLM-based arithmetic coding significantly outperforms traditional methods.  Moreover, a scaling law is observed in the compression performance, which somewhat corroborates the findings of \cite{huang2024compression}. 

    The experimental results presented in Table \ref{Table:blocksize} further investigate the influence of token block size on performance.  It can be observed that at higher SNR levels, the performance generally improves as the block size increases, indicating that larger block sizes facilitate enhanced semantic preservation, due to their ability to capture more contextual information.  However, at lower SNR levels, the performance declines with an increase in block size, suggesting that smaller blocks may be more resilient to avoid cumulative source decoding errors in these challenging scenarios.

    \begin{table*}[thp]
    \centering
    \caption{Influence of token block size on system performance during LLM-based arithmetic source encoding, for $\text{SNR}=\{-6,0,6\}$.} \label{Table:blocksize}
    \begin{tabular*}{.825\textwidth}{cccccccccccc}
    	\toprule[0.75pt]
    	\multirow{2}{*}[-2.6ex]{Block size}   & \multicolumn{3}{c}{Similarity} &  & \multicolumn{3}{c}{BLEU (1-gram)} &  & \multicolumn{3}{c}{BLEU (4-gram)}  \\
    	\cmidrule[0.36pt]{2-4}\cmidrule[0.36pt]{6-8}\cmidrule[0.36pt]{10-12}  & $-6$ & $0$ & $6$ &  & $-6$ & $0$ & $6$ &  & $-6$ & $0$ & $6$    \\
    	\midrule[0.5pt]
    	$16$            & \bm{$0.7708$} & \underline{$0.9157$} & $0.9993$ &   & \bm{$0.1975$} & \bm{$0.6452$} & $0.9877$ &   & \bm{$0.0072$} & \bm{$0.5081$} & $0.9830$   \\
    	$32$            & $0.7123$ & \bm{$0.9359$} & $0.9984$ &   & $0.1725$ & \underline{$0.5842$} & $0.9787$ &   & \underline{$0.0055$} & \underline{$0.4666$} & $0.9694$   \\
    	$64$            & $0.7001$ & $0.8938$ & \underline{$0.9999$} &   & $0.1160$ & $0.5801$ & \underline{$0.9969$} &   & $0.0018$ & $0.4270$ & \underline{$0.9922$}   \\
            $128$           & \underline{$0.7587$} & $0.8573$ & \bm{$ 0.9999$} &   & \underline{$0.1831$} & $0.4344$ & \bm{$0.9999$} &   & $0.0038$ & $0.2529$ & \bm{$0.9999$}   \\
    	\bottomrule[0.75pt]
    \end{tabular*}
    \end{table*}

\section{Conclusions and Discussions}
\label{sec:conclusion}
    In this paper, we present a comprehensive analysis and evaluation of SSCC, with a comprehensive comparison to JSCC in the context of SemCom. Our proposed SSCC framework, which integrates LLMs for source coding and ECCT for enhanced channel coding, demonstrates significant performance improvements over JSCC in terms of recovery performance at both word and semantic levels under both AWGN and Rayleigh fading channels, highlighting the potential effectiveness of SSCC in information transmission. 
    In particular, through extensive experiments, we validate the strong compressive capability of LLMs to eliminate redundancy in text and the robustness of ECCT in enhancing decoding reliability under various channel conditions. In a word, separate source channel coding is still what we need!
    
    Nevertheless, despite the validated performance superiority of SSCC, there remain several important issues worthy of further clarification and investigation.
    \begin{itemize}
    \item The performance evaluation of text transmission sounds inspiring. The proposed SSCC framework is channel-agnostic, while given the well-known generality issues, the DNN-based JSCC faces a performance decline when the channel changes significantly. 
    However, an extension to image transmission can be more challenging, and several issues like sequential tokenization require effective solutions. In this regard, potential solutions can incorporate 
    patch division from Vision Transformer (ViT) \cite{dosovitskiy2021an} to replace text tokenization, thereby segmenting images into semantic units for encoding. Consequently, the LLM-AC text predictor can be transformed into a probability modeler for image patches. Furthermore, the iterative decoding of ECCT can mitigate the error propagation issues in traditional JSCC, which is particularly crucial for multimedia transmission with high-fidelity requirements. 
    However, given the necessity for sequential prediction of image patches and the establishment of a structured directory system for their organization, the extension of the proposed framework to other modalities inherently poses significant technical challenges. Future research endeavors will be strategically directed toward resolving this critical issue, with the ultimate objective of achieving seamless integration between multimodal LLM and the SSCC architecture.
    On the other hand, our experimental experience indicates the accuracy of channel coding is of vital importance for end-to-end performance. Hence, we only consider a relatively low, fixed code rate here. However, systematic tuning of the code rate is also a worthwhile direction for future research. 
    \item There is no doubt that the involvement of LLMs in the SSCC framework requires substantial computational resources for both encoding and decoding processes. But we boldly argue that such involvement of pre-trained LLMs mainly is attributed to the underlying astonishing generalization capabilities that can handle a broad range of natural language datasets. Similar to JSCC methods, which often rely on training with specific datasets, we can also explore the feasibility of a much smaller model for typical datasets. In this way, we can substantially reduce computational overhead while maintaining reasonable performance. Corresponding experimental results for a Transformer model with significantly fewer parameters are provided in Appendix \ref{app:distillation}.
    \item This paper only considers the classical JSCC design, while ignoring the latest quantization and digital modulation techniques that have emerged in the development of JSCC. For example, \cite{tong_alternate_2023,tong_alternate_2024} show that utilizing a sparsity module to quantize the image embedding can yield a significant performance gain. However, \cite{tong_alternate_2023,tong_alternate_2024} have not compared these approaches with the astonishing capability of LLMs, and thus it remains unclear whether these amendments would enable JSCC to surpass LLM-based SSCC under a fair comparison. Nevertheless, given the inspiring results in this paper, there is no doubt that SSCC should be carefully improved rather than discarded.
    \item The discussions on JSCC are limited to the scenario to recover the semantics as accurately as possible. For SemCom \cite{lu_semanticsempowered_2024}, effectiveness-level or pragmatic communications may target at accomplishing different tasks under remotely controlled, noisy environments, rather than simple recovery of accurate semantics. In such cases, the underlying philosophy behind JSCC may promise independent merits.
    \item Extensive works have been conducted to improve the performance of model-free decoders. For example,  \cite{park2023mask} proposes a systematic and double mask eliminating the difficulty of identifying the optimal parity-check matrix (PCM) from numerous candidates from the same code. For performance enhancement on moderate code-length decoding, U-ECCT is proposed in \cite{nguyen2024u} inspired by U-Net, while in \cite{choukroun2022denoising}, the Denoising Diffusion Probabilistic Model (DDPM) \cite{ho2020denoising} is employed to model the transmission over channels as a diffusion process. Furthermore, a foundation model for channel codes is proposed in \cite{choukroun2024foundation} for application to unseen codes. Therefore, these recent works are worthy of evaluation in the SSCC framework.  
    \end{itemize}

\bibliographystyle{IEEEtran}
\bibliography{reference}

\appendices
\section{Pseudo Code of Finite Precision Arithmetic Codec}
\label{app:code}

\subsection{Pseudo Code for Encoder}
    Please refer to Algorithm \ref{alg:Finite-Precision Arithmetic Encoding}.
    \begin{algorithm}[!ht]
            \caption{Finite-precision arithmetic encoding.}
            \label{alg:Finite-Precision Arithmetic Encoding}
            \begin{algorithmic}[1]
                \REQUIRE {$N_k$: Current amounts of emitted bits $\bm{m}_{N_k}$}
                \REQUIRE {$p_{\text{cum}}(D_i|\bm{t}_{1:k})$: Cumulative probability of token $t_{k+1}=D_i \in \mathcal{D}$ given the first $k$ tokens}
                \REQUIRE {$l_k, u_k$: Current interval determined by the first k tokens}
                \REQUIRE {$\varepsilon_k$: Number of scaling bits}
                \STATE {\textbf{Initialization:}}
                \STATE {$N_{k+1} \gets N_k$}
                \STATE {$l_{k+1} \gets l_k + (u_k - l_k) p_{\text{cum}}(D_{i-1} | \bm{t}_{1:k})$ \algocomment{If $k=0$, use $p_{\text{cum}}(D_{i-1})$ }}
                \STATE {$h_{k+1} \gets l_k + (u_k - l_k) p_{\text{cum}}(D_i | \bm{t}_{1:k})$ \algocomment{If $k=0$, use $p_{\text{cum}}(D_i)$}} \vspace{-\baselineskip}
                \STATE {$\varepsilon_{k+1} \gets \varepsilon_k$}
                \STATE {\textbf{Scaling:}}
                \WHILE{any of the scaling conditions is met}
                    \IF{$u_{k+1} < 0.5$}
                        \STATE {\algocomment {Scaling 1}}
                        \STATE {$l_{k+1}, u_{k+1} \gets 2l_{k+1}, 2u_{k+1}$}
                        \STATE {$\bm{m}_{N_{k+1}+1} \gets 0$ \algocomment{Emit one bit's $0$}}
                        \STATE {$\bm{m}_{N_{k+1}+2:N_{k+1}+1+\varepsilon_{k+1} } \gets 1$ \algocomment{Emit $\varepsilon_{k+1}$ bits' $1$}}
                        \STATE {$N_{k+1} \gets N_{k+1} + 1 + \varepsilon_{k+1}$}
                        \STATE {$\varepsilon_{k+1} \gets 0$}
                    \ELSIF{$l_{k+1} \ge 0.5$}
                        \STATE {\algocomment {Scaling 2}}
                        \STATE {$l_{k+1}, u_{k+1} \gets 2(l_{k+1} - 0.5), 2(u_{k+1} - 0.5)$}
                        \STATE {$\bm{m}_{N_{k+1}+1} \gets 1$ \algocomment{Emit one bit's $1$}}
                        \STATE {$\bm{m}_{N_{k+1}+2:N_{k+1}+1+\varepsilon_{k+1} } \gets 0$ \algocomment{Emit $\varepsilon_{k+1}$ bits' $0$}}
                        \STATE {$N_{k+1} \gets N_{k+1} + 1 + \varepsilon_{k+1}$}
                        \STATE {$\varepsilon_{k+1} \gets 0$}
                    \ELSIF{$0.25 \le l_{k+1} < 0.5 \le u_{k+1} < 0.75$}
                        \STATE {\algocomment {Scaling 3}}
                        \STATE {$l_{k+1}, u_{k+1} \gets 2(l_{k+1} - 0.25), 2(u_{k+1} - 0.25)$}
                        \STATE {$\varepsilon_{k+1} \gets \varepsilon_{k+1} + 1$}
                    \ENDIF
                \ENDWHILE
                \RETURN $N_{k+1}$, $\bm{m}_{N_{k}+1:N_{k+1}}$ \algocomment{Updated emitted bits}
            \end{algorithmic}
        \end{algorithm}
\subsection{Pseudo Code for Decoder}
Please refer to Algorithm \ref{alg:Finite-Precision Arithmetic Decoding}.
    \begin{algorithm}[!ht]
    \caption{Finite-precision arithmetic decoding.}
    \label{alg:Finite-Precision Arithmetic Decoding}
    \begin{algorithmic}[1]
        \REQUIRE {$K_n$: Current amounts of decoded tokens}
        \REQUIRE {$p_{\text{cum}}(D_i|\bm{t}_{1:K_n})$: Cumulative probability of token $t_{K_{n+1}+1} = D_i\in \mathcal{D}$ given the first $K_n$ tokens}
        \REQUIRE {$l_n, u_n$: Current interval determined by the first $n$ bits}
        \REQUIRE {$l_{K_n}, u_{K_n}$: Interval of sequence $\bm{t}_{1:K_n}$ that has been decoded}

        \STATE {\textbf{Initialization:}}
        \STATE {$K_{n+1} \gets K_n$}
        \STATE {$l_{K_{n+1}}, h_{K_{n+1}} \gets l_{K_n}, h_{K_n}$}
        \IF{the $(n+1)$-th bit $m_{n+1} = 0$}
            \STATE {$l_{n+1}, h_{n+1} \gets l_n, \frac{1}{2}(l_n+h_n)$}
        \ELSE
            \STATE {$l_{n+1}, h_{n+1} \gets \frac{1}{2}(l_n+h_n), h_n$}
        \ENDIF

        \WHILE{Not End-of-Sentence symbol} 
            \STATE {\textbf{Search:}}
            \STATE {Find $D_i \in \mathcal{D}$ such that:}
            \STATE \hspace{1em}{$L = l_{K_{n+1}} + (u_{K_{n+1}} - l_{K_{n+1}}) p_{\text{cum}}(D_{i-1} | \bm{t}_{1:K_n})$ \algocomment{If $K_n=0$, use $p_{\text{cum}}(D_{i-1})$}}
            \STATE \hspace{1em}{$U = l_{K_{n+1}} + (u_{K_{n+1}} - l_{K_{n+1}}) p_{\text{cum}}(D_{i} | \bm{t}_{1:K_n})$ \algocomment{If $K_n=0$, use $p_{\text{cum}}(D_{i})$}}
            \STATE \hspace{1em}{$L \le l_{n+1} < u_{n+1} < U$ \algocomment{i.e. current interval of the $n$-th bit is included in the interval of $D_i$}}

            \IF{$D_i$ exists}
                \STATE {\textbf{Update:}}
                \STATE {$K_{n+1} \gets K_{n+1} + 1$}
                \STATE {$t_{K_{n+1}} \gets D_i$ \algocomment{Output $D_i$ to the token sequence $\bm{t}$}}
                \STATE {$l_{K_{n+1}}, u_{K_{n+1}} \gets L, U$}

                \STATE {\textbf{Scaling:} Similar to the Scaling in Algorithm \ref{alg:Finite-Precision Arithmetic Encoding}}
                \STATE {Go to \textbf{Search}}
            \ELSE
                \RETURN $K_{n+1}$, $\bm{t}_{K_{n}+1:K_{n+1}}$
            \ENDIF
        \ENDWHILE

        \RETURN $K_{n+1}$, $\bm{t}_{K_{n}+1:K_{n+1}}$ \algocomment{Updated decoded tokens}
    \end{algorithmic}
\end{algorithm}

\section{Key Modules of ECCT}
\label{app:ecct}
\subsection{Positional Reliability Encoding}
For the channel output $\bm{y}$, the positional reliability encoding transforms each dimension of $\bm{\Tilde{y}}$ into a high \emph{d} dimensional embedding $\bm{\phi}$, which enriches the information of input embedding vectors and replaces $\bm{\Tilde{y}}$ as the input of ECCT, defined by
\begin{equation}
    \bm{\phi}_i = 
    \begin{cases}
            |\bm{y}_i|\bm{W}_i, &  \text{if \space} i \leq N; \\
            \binsign\big(\syn(\bm{y}_{i-N+1})\big)\bm{W}_i, & \text{otherwise}.
    \end{cases}
\end{equation}
where $\{ \bm{W}_i \in \mathbb{R^d}\}_{i=1}^{2N-K}$ denotes the learnable embedding matrix representing the bit's position dependent one-hot encoding. The encoding method corresponds to the input reliability and is positional, since unreliable information of low magnitude would collapse to the origin, while the syndrome would scale negatively, hence the name positional reliability encoding.

\begin{figure*}[tb]
        \centering
        \includegraphics[width=0.815\textwidth]{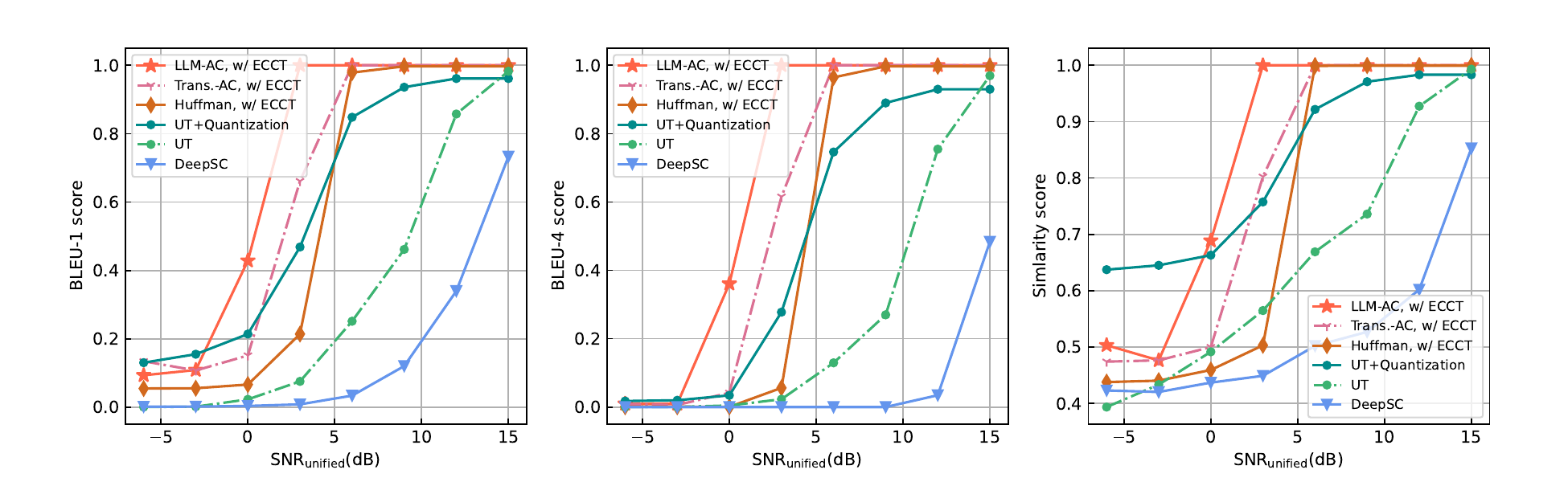}
        \caption{BLEU and Similarity scores versus $\textbf{SNR}_{\textbf{unified}}$ are evaluated for the same number of transmitted symbols. Compared to Fig. \ref{fig:Comparison_AWGN}, a transformer trained on the European Parliament dataset, which replaces the exact role of the LLM, is added for comparison.}
        \label{fig:Comparison_t_AWGN}
    \end{figure*}
    \begin{figure*}[tb]
        \centering
        \includegraphics[width=0.815\textwidth]{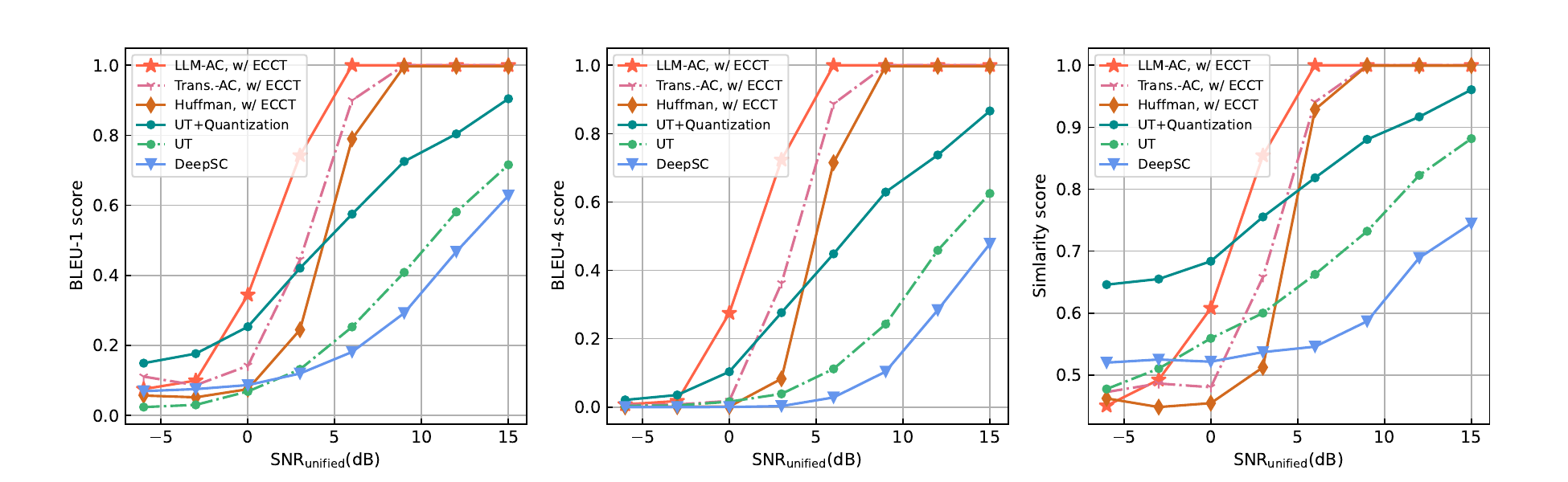}
        \caption{BLEU and Similarity scores versus $\textbf{SNR}_{\textbf{unified}}$ for the same number of transmitted symbols. Compared to Fig. \ref{fig:Comparison_Rayleigh}, a transformer trained on the European Parliament dataset, which replaces the exact role of the LLM, is added for comparison.}
        \label{fig:Comparison_t_Rayleigh}
    \end{figure*}

\subsection{Code Aware Self-Attention}

\begin{algorithm}[!t]
    \caption{Pseudo code of building the attention mask.}
    \label{alg:building attention mask}
    \begin{algorithmic}[1]
        \REQUIRE {parity-check matrix $\bm{H}$ of error correction code $C_e(N,K)$}
        \STATE{$\text{mask} \gets \texttt{eye}(2N-K)$}
        \FOR{$i = 1,2,...,N-K$}
            \STATE{$\text{idx} \gets \text{where}(\bm{H}[i] == 1)$}
            \FOR{$j$ in idx}
                \STATE{$\text{mask}[N+i, \text{\space} j], \text{\space} \text{mask}[j, \text{\space} N+i] \gets 1$}
                \FOR{$l$ in idx}
                    \STATE{$\text{mask}[j, \text{\space} l], \text{\space} \text{mask}[l, \text{\space} j] \gets 1$}
                \ENDFOR
            \ENDFOR
        \ENDFOR
        \STATE{$\text{mask} \gets -\infty(\neg \text{\space mask})$}
        \RETURN mask \algocomment{Output attention mask $\bm{g(H)}$}
    \end{algorithmic}
\end{algorithm}

The code aware attention mask mechanism aims to integrate code-specific sparse marks which incorporate the inherent structural characteristics of their respective PCM as the domain knowledge. Given a codeword defined by generator matrix $\bm{G}$ and parity check matrix $\bm{H}$, the attention mask is defined by $ \bm{g(H)} : \{ 0,1 \}^{(N-K) \times N} \rightarrow \{ -\infty,0 \}^{(2N-K)\times(2N-K)}$, the construction of which is shown in Algorithm \ref{alg:building attention mask}. Then the code aware self-attention mechanism could be represented as
\begin{equation}
    A_H(\bm{Q,K,V}) = \text{Softmax}\left(\frac{\bm{QK}^T+\bm{g(H)}}{\sqrt{d}}\right)\bm{V}.
\end{equation}
where $\bm{Q}$, $\bm{K}$\footnote{It is worthwhile to point that $\bm{K}$ distinguishes from $K$, which represents the length of the error correction code in the previous text.} and $\bm{V}$ denote the query, key and value in self-attention.
During the implementation, the code aware attention mask mechanism is used as an enhancement of the multi-head-attention module in the classical transformer architecture.

\section{Experimental Results of Model Distillation}
\label{app:distillation}
In this part, we utilize a Transformer model with significantly fewer parameters while retaining the LLM's tokenizer and performing self-supervised training on the dataset of the European Parliament \cite{koehn2005europarl}. Correspondingly, on top of Fig. \ref{fig:Comparison_AWGN} and Fig. \ref{fig:Comparison_Rayleigh}, a transformer trained on the European Parliament dataset, which replaces the exact role of the LLM, is added for comparison. 
The related results in Fig. \ref{fig:Comparison_t_AWGN} and Fig. \ref{fig:Comparison_t_Rayleigh} indicate consistent observations. Moreover, Table \ref{Table:flops} compares the utilized parameters and FLOPs in the aforementioned frameworks, indicating the feasibility and superiority of SSCC under comparable computation overhead.

\begin{table}[h]
\centering
\caption{Comparison of parameters and FLOPs between different systems.} \label{Table:flops}
\begin{tabular}{ccc}
    \toprule[0.75pt]
    Model    &    Params    &    FLOPs    \\
    \midrule[0.5pt]
    LLM-SSCC       &    $124.51M$    &    $\approx960G$ \\
    Trans.-SSCC    &    $0.33M$    &    $\approx4.5G$ \\
    DeepSC         &    $10.58M$    &    $\approx57G$ \\
    UT             &    $18.43M$    &    $\approx12G$ \\
    \bottomrule[0.75pt]
\end{tabular}
\end{table}

\end{document}